\documentclass[sigconf]{acmart}

\usepackage{booktabs} 
\usepackage{listings}
\usepackage{paralist}

\usepackage{incgraph,tikz}
\usepackage{graphicx}
\usepackage{longtable}
\usepackage{wrapfig}
\usepackage{subcaption}
\usepackage{hyperref}
\usepackage{multirow} 
\usepackage{balance}

\usepackage{csquotes}

\usepackage{pifont}
\newcommand{\xmark}{\ding{55}}%
\usepackage{xcolor,colortbl}
\newcommand{\floor}[1]{$\lfloor$#1$\rfloor$}
\usepackage{enumitem}

\copyrightyear{2024}
\acmYear{2024}
\setcopyright{acmlicensed}\acmConference[CODASPY '24]{Proceedings of the Fourteenth ACM Conference on Data and Application Security and Privacy}{June 19--21, 2024}{Porto, Portugal}
\acmBooktitle{Proceedings of the Fourteenth ACM Conference on Data and Application Security and Privacy (CODASPY '24), June 19--21, 2024, Porto, Portugal}
\acmDOI{10.1145/3626232.3653276}
\acmISBN{979-8-4007-0421-5/24/06}

\begin{document}

\title{Stop Stealing My Data: Sanitizing Stego Channels \\ in 3D Printing Design Files}

\author{Aleksandr Dolgavin}
\affiliation{
    \institution{Auburn University}
    \country{USA}
}
\author{Mark Yampolskiy}
\affiliation{
    \institution{Auburn University}
    \country{USA}
}
\author{Moti Yung}
\affiliation{
    \institution{Google LLC, Columbia University}
    \country{USA}
}

\begin{abstract}

The increased adoption of additive manufacturing (AM) and the acceptance of AM outsourcing created an ecosystem in which the sending and receiving of digital designs by different actors became normal. It has recently been shown that the STL design files---most commonly used in AM---contain steganographic channels. 
Such channels can allow additional data to be embedded within the STL files without changing the printed model. These factors create a threat of misusing the design files as a covert communication channel to either exfiltrate stolen sensitive digital data from organizations or infiltrate malicious software into a secure environment. This paper addresses this security threat by designing and evaluating a \emph{sanitizer} that erases hidden content where steganographic channels might exist. The proposed sanitizer takes into account a set of specific constraints imposed by the application domain, such as not affecting the ability to manufacture part of the required quality using the sanitized design.

\end{abstract}



\renewcommand{\shortauthors}{Dolgavin et al.}
\renewcommand{\shorttitle}{Stop Stealing My Data}

\maketitle

\section{Introduction}

There are several driving factors for this trend. First, just-in-time and on-demand manufacturing brings down the need for storage and transportation, thus reducing the overall expenditures. 
Second,  redesigning existing parts for AM allows part consolidation, reducing the need for assembly while improving the system's reliability. 
Last but not least, generative design for AM produces function-optimized part geometries, thus decreasing overall weight, contributing to the fuel efficiency of systems employing AM parts.

A digital design file defines the geometry of a part manufactured by an AM machine. Universality of AM machines and the reliance on design files make it possible to manufacture parts on different machines. At the same time, the costs of industrial-grade AM machines can be prohibitive for many organizations, e.g., industrial-grade metal AM machines often cost over \$500k. This creates the opportunity and  motivation for new business models when various AM companies provide manufacturing as a service. There is a wide acceptance among manufacturers for providing and using AM services~\cite{yampolskiy2022state}.
This is reflected by the fact that more than 1200 companies worldwide provide AM as a service as of 2024~\cite{3ddirectory2024website}. 

In the AM-as-a-Service business model, design files are commonly exchanged between customers and manufacturing companies known as \emph{AM service providers} or \emph{service bureaus}, thus crossing the organizational domain boundaries. In addition, numerous content sharing websites such as Thingiverse~\cite{thingiverse2023website} offer digital design files for download, either for free or for a small fee. Such sites are often used by hobbyists who own desktop 3D printers. This creates an environment where downloading a design file from the Internet or sending it to a third company becomes common. Furthermore, to send, receive, and download digital design files for AM, a variety of protocols are used, including HTTP/HTTPS (web) or even POP3/IMAP (e-mail). Thus, such activities do not automatically raise any ``red flags'', creating an environment conducive to using such files as ``carriers'' of steganographically embedded information. 

Yampolskiy et al.~\cite{yampolskiy2021what} have demonstrated several use cases of such misuse:
design files sent outside of a secure environment could contain stolen sensitive information; design files downloaded into a secure environment could contain malicious software; design files can also be used as a storage medium for hiding illegal or unlawfully possessed data. 
All these constitute serious security risks that need to be addressed.
However, neither contemporary firewalls nor antivirus scanners can even recognize let alone prevent such activities. 

This paper addresses this threat for the STL design file format. While there are multiple design file formats for AM, STL (STereoLithography) remains the most widely adopted across the industry~\cite{asm2020handbook24a}. For this file format, we design, develop, and evaluate a \emph{sanitizer} also known as \emph{Content Threat Removal} (CTR)~\cite{wiseman20175}  that eliminates any information that could have been embedded in previously discovered STL stego channels. Once an STL file has been sanitized, it can no longer contain unauthorized or malicious payload. While doing this, we ensure that neither the 3D geometry described in an STL file nor the ability to 3D print it is affected.

This paper is structured as follows. In Section~\ref{sec:background} we outline the STL file format and the stego channels that have been previously discovered in it. In Section~\ref{sec:challenges} we analyze the characteristics of these channels and the resulting implications (that is, requirements and constraints) on the STL sanitizer. 
In Section~\ref{sec:sanitizing} we present the developed sanitizer that satisfies the identified requirements. 
We present the conducted evaluation of the implemented sanitizer in Section~\ref{sec:evaluation}. 
A discussion of sanitizer characteristics is presented in Section~\ref{sec:discussion}. 
We conclude this paper with a brief outline of the planned follow-up research.

\section{Background \& Related Work}
\label{sec:background}

\subsection{STereoLithography (STL) File Format}
\label{sec:related_work:STL_file_format}

STereoLithography (STL) is a legacy file format originally developed for subtractive manufacturing and later adopted in additive manufacturing.
While there are several more modern and potent alternative file formats, most prominently AMF and 3MF, STL remains the most widely adopted in hobbyist and industrial settings~\cite{asm2020handbook24a}.

An STL describes a 3D object as a collection of triangular \emph{facets} that enclose the ``water tight'' 3D object's surface. Each such facet is uniquely defined by its three \emph{vertices} ${(v_1, v_2, v3)}$, each of which, in turn, is defined by its {(x, y, z)} coordinates. In addition, every facet is associated with a \emph{normal} vector, which determines which side of the facet is outwards facing. It is expected that the order in which the vertices of a facet are listed and the normal vector follow the ``right-hand rule,'' which is defined similarly to the one in electromagnetism. Beyond this, no other restrictions are imposed on the STL elements.

\begin{figure}[tbp]
    \centering
    \begin{subfigure}[b]{\linewidth}
        \centering
		\includegraphics[width=\textwidth]{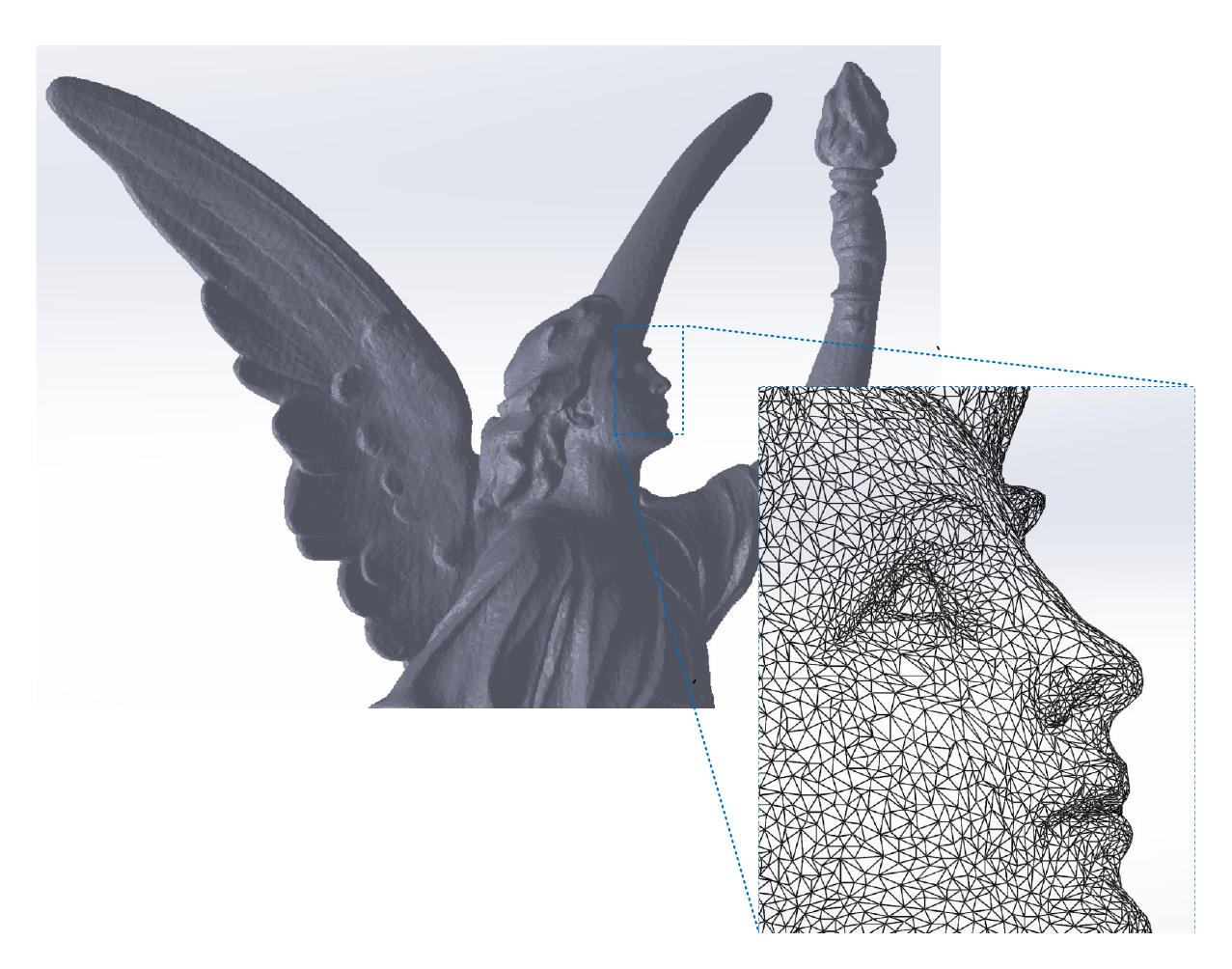}
        \caption{Solid and Wired (Triangular Facets) Representations}
		\label{fig:STL_example_3DView}
    \end{subfigure}%
    
~

    \begin{subfigure}[b]{\linewidth}
    \centering
    \includegraphics[width=\textwidth]{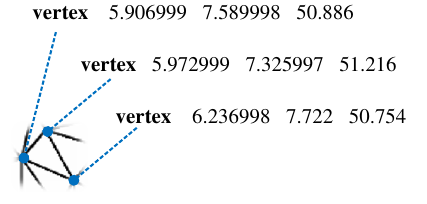}
    \caption{3D Facet Defined by its Three Vertices}
    \label{fig:STL_SingleFacet_3D}
    \end{subfigure}

~

    \begin{subfigure}[b]{\linewidth}
        \centering
	    \begin{lstlisting}[backgroundcolor = \color{lightgray!10}, frame = single] %language = C,
                   
solid StanfordLucy
  facet normal -0.1128 -0.818 -0.5641
    outer loop
      vertex -13.101 0.527998 52.206
      vertex -13.035 0.791999 51.81
      vertex -12.771 0.527998 52.14
    endloop
  endfacet
  facet normal -0.0573 0.774 0.6306
    outer loop
      vertex 5.906999 7.589998 50.886
      vertex 5.972999 7.325997 51.216
      vertex 6.236998 7.722    50.754
    endloop
  endfacet
	...
endsolid StanfordLucy
        \end{lstlisting}
        \caption{ASCII STL File (Excerpt: Two Triangular Facets Shown)}
		\label{fig:STL_example_ASCII}
    \end{subfigure}

	\caption{Description of the 3D digital design model as STereoLithography (STL) file format~\cite{yampolskiy2021what}.}
	\label{fig:STL_example}
\end{figure}

An STL file can have both human-readable ASCII and more space-efficient binary representation, which are semantically equivalent. The ASCII representation is depicted in Figure~\ref{fig:STL_example}. 
A single STL file can contain descriptions of several ``solid'' objects.
In STL, the keyword \emph{solid} introduces the entire 3D object; 
a pair of \emph{facet} and \emph{endphacet} keywords enclose the description of a single facet; it is followed by \emph{normal} associated with the entire facet;
within the facet and enclosed in a pair of \emph{outer loop} and \emph{endloop} keywords, three individual vertices are defined, each of which is introduced with the keyword \emph{vertex}. Individual values associated with x, y, and z can be defined in a scientific or standard notation.

\subsection{Stego Channels in STL}
\label{sec:related_work:STL_stego_channels}

Under the definition described above, the object description contains several sources of entropy whose manipulation would not affect the defined STL geometry or its printability~\cite{yampolskiy2021what, yampolskiy2022crypto}. First, the order in which the facets are listed in the STL file is unimportant because it does not change the overall geometry described. This constitutes \emph{facet} stego channel. Second, the same applies to the order in which vertices are listed in the definition of a facet, as long as they remain adherent to the right-hand rule. This constitutes \emph{vertex} stego channel. These two stego channels are always present in STL files, regardless of whether the STL file is in ASCII or binary format.

The third stego channel is only present in ASCII STL files. STL allows for the use of both scientific and normal notation to represent numbers. 
The notation in which individual values are defined or the location of the decimal point does not affect the value (if numerical errors are within limits that can be ignored). Because the same file can contain a mixture of notations for various numbers, this constitutes the \emph{number representation} stego channel.

Lastly, some \emph{slicers}\footnote{Slicer is a program that translates the description of 3D geometry into a sequence of adjacent layers that must be printed upon each other. For each of these layers and layer transitions, the slicer generates a sequence of printer commands, known as a \emph{toolpath}. There are numerous toolpath formats, including proprietary for individual OEMs. An example of a toolpath format is \emph{G-Code}, which is frequently utilized by consumer-grade desktop 3D printers.} tend to ignore the normal vector defined in STL and instead recalculate it based on the vertices in a facet~\cite{yampolskiy2022crypto}.
With such slicers, violation of the right-hand rule (RHR) by the normal vector will neither affect the design's printability nor the produced object's geometry.
In such cases, adherence or violation of the normal vector to the RHR constitutes \emph{normal} stego channel.

While the latter two described stego channels only conditionally possible, the first two are always present in ASCII and binary STL files. In addition, in ASCII STL files, all kinds of classical stego channels are possible, such as insertion of space and tabulators, use of comments, etc.

All the above-mentioned stego channels represent a transmission medium; an encoding scheme determines how specific variations in this medium translate into bits of information. Different encoding schemes can be defined to determine what represents the bit value 0 or 1 (in a binary encoding) or other set of values (in the case of a higher base encoding scheme). For example, a normal stego channel can be defined as follows: adherence to the right-hand rule can correspond to the bit value 0, and violation of the rule to the bit value 1. However, in general, the multitude of possible encoding strategies poses one of the challenges for the sanitizer; as such, we will discuss it to a greater extent in the following Section~\ref{sec:challenges:encoding-multitude}

\begin{table}
    \centering
    \caption{Stego channels in STL file format. For each channel, the following properties are specified: availability for ASCII and Binary STL representation (\checkmark - available, (\checkmark) - conditionally available, \xmark - unavailable), and a conservative (floor) estimate of the possible alternative encoding strategies.}
    
    \begin{tabular}{|l|c|c|c|}
        \hline
        \textbf{\textsc{Channel}} & \textbf{\textsc{ASCII}} & \textbf{\textsc{Bin.}} & \textbf{\textsc{\floor{\# Enc.}}} \\ 
        \hline
        \hline
        \emph{Facet} & \checkmark & \checkmark & 6M!+ \\
        \hline
        \emph{Vertex} & \checkmark & \checkmark & 90+ \\
        \hline
        \emph{Number} representation & \checkmark & \xmark & 2+ \\
        \hline
        \emph{Normal} vector adherence to RHR & (\checkmark) & (\checkmark) & 2+ \\
        \hline
        \emph{Classical ASCII} stego channels & \checkmark & \xmark & 2+ \\
        \hline
    \end{tabular}
    
    \label{tab:stego-channels}
\end{table}

\subsection{Related Work}
\label{sec:related_work}

The only two prior works specifically on stego channels in digital design files used in AM were published by Yampolskiy et al.~\cite{yampolskiy2021what, yampolskiy2022crypto}. 
In their original work~\cite{yampolskiy2021what}, authors discover the sources of entropy in this design file format, show how to use these for the encoding of arbitrary information, and discuss how they can be used for various nefarious activities such as data in- and exfiltration. 
In the second~\cite{yampolskiy2022crypto}, authors present a method for how the same channels can be used for design watermarking and fingerprinting, enabling assertion of digital design ownership rights and identification of an actor leaking the design, respectively.

For the sake of completeness, we should note that multiple physical object watermarking techniques have been developed specifically for 3D printing~\cite{wang2008comprehensive, macq2015applicability, yampolskiy2018security}. 
These generally embed a bit stream through the imperceptible (by a human eye) distortions of the printed object geometry~\cite{ohbuchi1998watermarking, ohbuchi1997watermarking, kanai1998digital, praun1999robust, ohbuchi2001watermarking, cayre2003data, hou20153d, delmotte2019blind} or selective material substitution~\cite{silapasuphakornwong20193d}.
Use of 3D-printed barcodes or QR codes has also been proposed, printed either on or under the object's surface~\cite{chen_embedding_2019}.
By their nature, all these techniques affect one or multiple of the part's three F's, Fit, Form, and Function~\cite{gibson2014additive}.
However, such changes make applications of this kind of watermark to functional parts problematic~\cite{yampolskiy2022state}.
As shown by numerous intentional AM sabotage attacks, geometric deviations can introduce flaws in the part that could cause its premature failure~\cite{sturm2014cyber, belikovetsky2016dr0wned, graves2021sabotaging}.

We are unaware of prior works on sanitizing hidden information in additive manufacturing design files. 
However, there is a substantial body of literature on both steganographic channels and sanitizing information in other data formats.
For example, stego channels have been found and used in data formats like BMP~\cite{nozaki1998large}, JPEG~\cite{skodras2001jpeg}, GIF~\cite{tiwari2010evaluation}.
To encode information, various strategies have been utilized, including spatial methods such as the classical Least Significant Bit (LSB)~\cite{johnson1998exploring}, frequency (also called transform) ~\cite{khayam2003discrete} and batch steganography~\cite{ker2006batch}.
Steganalysis aims to answer whether a secret message has been embedded in one of the covert channels. 
Classical statistic approaches~\cite{westfeld1999attacks, sabeti2007steganalysis, chen2018binary} have been meanwhile augmented with machine learning-based and pooled steganalysis~\cite{boroumand2017applications, ma2018selection, cogranne2017practical}.
Orthogonal to the question of distinguishability, the need for stego channel sanitization has been recognized for a variety of application scenarios, including transmitted digital images~\cite{corley2019destruction}, protection of cryptographic primitives against kleptographic
attacks~\cite{russell2016destroying}, online social networks~\cite{zhu2021destroying}, and access control~\cite{wang2021cross}, to name a few.
We point an interested reader to a recent survey by Muralidharan et al.~\cite{muralidharan2022infinite} that provides a comprehensive overview of steganography approaches and countermeasures.


\section{Analysis: Characteristics of STL Steganographic Channels  \& their Implications on Sanitizing}
\label{sec:challenges}

We analyzed and identified several differences between classical stego channels considered in cybersecurity and those in STL files.
These differences pose challenges and restrict the methods available for sanitizing various STL stego channels. At the end of each challenge-specific discussion, we briefly explain its consequences for the sanitizing process.

\subsection{Multitude of Encoding Strategies}
\label{sec:challenges:encoding-multitude}

Table~\ref{tab:stego-channels} summarizes STL stego channels and their presence in this file format's ASCII or binary representation. It also specifies a conservative estimate of the possible alternative encoding strategies that we will focus on in this section. To proceed successively from simpler to more complex cases, we will discuss the channels in the order inverse to the one used in the table.
 
The classical ASCII channel is not a part of the geometry description. 
For example, a simplest encoding strategy could use tabulator and spaces (both of which separate individual elements in ASCII STL) to represent either the bit value 0 or 1.
Such channels can be easily ``erased'' by loading and simply uniformly re-saving STL files. 

For the normal stego channel, generalizing upon the example we outlined at the end of the prior section, adherence to the right-hand rule (RHR) can correspond to either the bit value 0 or 1, while the violation of the rule to the complementary bit value, thus accounting for two different encodings. Note that more sophisticated encoding strategies are possible even without violation of RHR, e.g., by deviating the normal vector length or direction from the one that can be calculated based on the three vertices' coordinates.
Furthermore, these could represent more than a single bit value.

Similarly, standard and scientific number notations can correspond to bit values 0 and 1, or vice versa, 1 and 0. However, even if exclusively the scientific number notation is used, the location of the decimal point can be used to encode values at higher order bases, thus dramatically increasing the number of possible encoding assignments.

The encoding definition is generally more complex for the vertex and facet channels. Defining encoding on various STL stego channels requires a distinction between different cases to which 3D printing is oblivious. This requires an introduction of \emph{order}, which can be defined in numerous ways. 

For vertex stego channel, we need to compare between any two vertices a facet is comprised of. This is a successive comparison between their coordinates x, y, and z (see Figure~\ref{fig:code_maxvertex}). However, a malicious actor can select the order in which coordinates are compared differently, e.g., y, then x, and only then z. This results in 3! (=6) different order definitions, even if numerical values are compared. Further definitions are also possible, which would only partially lead to the same results, e.g., comparing two strings representing concatenated x-y-z coordinates in a standard notation.

\begin{figure}[h!]
    \centering
        \begin{lstlisting}[backgroundcolor = \color{lightgray!10}, frame = single, language = C] % , basicstyle=\tiny

func MaxVertex(v1, v2)
{
  if (v1.x < v2.x) return v2; else if (v1.x > v2.x) return v1
  
  if (v1.y < v2.y) return v2; else if (v1.y > v2.y) return v1
  
  if (v1.z < v2.z) return v2; else if (v1.z > v2.z) return v1;

  return v1;
}

        \end{lstlisting}
    
\caption{Possible definition of \emph{order} in vertex stego channel.}
	\label{fig:code_maxvertex}
\end{figure}

Upon establishing the order, bit encoding can be introduced. Because the vertex stego channel is a well-formed STL file, all vertices of a facet are distinct, and three distinct rotation positions are possible. Upon this set, either base-2 (binary) or base-3 encoding can be defined. 
For example, if ${v_1}$ listed first in the description of a facet is the biggest of the three, we can associate with this case bit value 1; otherwise, bit value 0: \\
 
\begin{math}
\boxed{
\textit{Vertex\_Base2\_EncodedBit} \text{ =} 
  \begin{cases}
    \text{1,} & \text{if } \textit{v}_\textit{1} \text{=} \textit{max(v}_\textit{1}\text{,  }\textit{max}\text{(}\textit{v}_\textit{2}\textbf{,} \textit{v}_\textit{3}\text{))}\\
    \text{0,} & \text{otherwise.}
  \end{cases} 
}
\end{math} \\

Thus, in base-2 encoding, we can distinguish between at least six variants to define bit values 0 and 1 based on the first vertice listed: whether it is the smallest, medium, or largest vertex that determines which bit value is assigned, with two different assignments (0 or 1) possible per each case. For base-3, nine different encoding variants can be defined based on which of the values (0, 1, and 2) is assigned to each case (smallest, medium, largest vertex). Even without considering more obscure encoding schemes (e.g., comparing the second of the vertices instead of the first one), it will yield 90 = (=6*(6+9)) different possible encoding schemes for the vertex stego channel.

For encoding on facet stego channel, the order and the number of involved facets impact the number of possible encodings. To establish order between any two facets ${f_i}$ and ${f_j}$ (note that these facets do not have to be subsequent!), their three vertices must be compared. The number of permutations in which they can be compared is 6 (= 3!), even without considering the multitude introduced by the necessity to introduce the order between vertices (as described above for the vertex stego channel). 

The number of facets used in encoding determines the maximal base. In the simplest case, when just two consecutive facets are used, they can only represent a single bit (i.e., base-2). In the case of three facets, up to base-6 is possible (= 3!), etc. Generalizing for M facets would result in base-M!. Note that this estimate does not consider the possibility of using non-neighbor facets or strategies that would make encoding robust to a simplistic sanitization.

\paragraph{\textbf{Consequences for Sanitizing:}} 
\textbf{[C1]} 
As the encoding selected by a malicious actor cannot be distinguished from the STL file alone, it cannot be either presumed or used in a sanitizer. This applies to all stego channels in STL.

\subsection{Orthogonality of STL Stego Channels \& Existence of Robust Encoding Strategies}
\label{sec:challenges:robust-encoding}

Two other characteristics of STL stego channels are somewhat intertwined. First, as identified in the original paper by Yampolskiy et al.~\cite{yampolskiy2021what}, the STL stego channels are orthogonal, i.e., data can be stored in and accessed from multiple channels simultaneously. Having said this, we discovered that these channels are asymmetrically independent; 
that is, disruption in one channel might or might not affect the information stored in another.

For example, both facet and vertex channels can be used at once, e.g., to store the same data redundantly. Both these data streams can be accessed independently. In this example, should only the vertex stego channel be sanitized, the information can still be obtained from the facet stego channel. 
That is presuming that the facet comparison is made after transforming their vertex channel to a \emph{canonical} form, as will be exemplified below.
One could assume that if the facet stego channel is sanitized instead, both channels would be disrupted. However, this is only conditionally true.

While working on the STL sanitizer, we identified a second property that needs to be considered: several encoding strategies that would be robust against simple sanitization efforts can be defined. Staying with the example above, a signaling protocol could be defined, in a \emph{codebook} manner, as a difference between the number of 0's and 1's encoded in the vertex stego channel across the entire STL file. Should exclusively facet channel be sanitized, while the location of individual 0's and 1's in the vertex channel will be perturbed, this would not affect the message encoded there.

\paragraph{\textbf{Consequences for Sanitizing:}} 
\textbf{[C2]} 
We cannot assume which STL stego channels have been used to encode information. Furthermore, sanitizing just a few selected STL stego channels might not be enough to erase the information stored in or across other channels. Therefore, all identified STL stego channels need to be sanitized.

Furthermore, we provide an additional example of a robust encoding strategy without attempting to be complete. In Yampolskiy et al.~\cite{yampolskiy2021what}, authors used two subsequent facets ${f_i}$ and ${f_{i+1}}$ to encode a single bit of information. Should sanitizers presume this encoding strategy and overwrite with random bits, an encoding strategy that is resistant to this approach could be defined. For example, it can be defined on two consecutive pairs of facets as follows: \\

\begin{math}
\boxed{
\textit{Facet\_EncodedBit} \text{ =} 
\begin{cases}
\text{1,} & \text{if } C_{facet}(C_{vert}(f_{i}), C_{vert}(f_{i+1})) > \\
          & C_{facet}(C_{vert}(f_{i+2}), C_{vert}(f_{i+3}))\\
\text{0,} & \text{otherwise.}
\end{cases} 
}
\end{math} \\

Here, ${C_{vert}}$ is a function transferring the vertex channel of a facet to its \emph{canonical} representation. This representation can be defined arbitrarily, e.g., by rotating vertices till the state when the first one is the smallest. Similarly, ${C_{facet}}$ is a \emph{canonical} form that can be defined on facet stego channel as the ``smallest facet first,'' automatically negating the effect of the sanitizer that assumes the encoding in the facets of i and i+1. Therefore, the information stored in two pairs of faces (${f_{i}}$, ${f_{i+1}}$) and (${f_{i+2}}$, ${f_{i+3}}$) would remain present even after naive sanitization.

\paragraph{\textbf{Consequences for Sanitizing:}} 
\textbf{[C3]} 
We cannot assume which elements of STL represent a single bit (or higher base information) within an STL stego channel. Thus, sanitizer should be agnostic to it.

\subsection{3D Printing Application Requirements: Printability and Print Quality}

Digital files containing stego channels are generally purpose-bound. Classically, image, audio, or video file formats are considered. 
For these cases, the modifications introduced in the stego channel (i) should not prevent the digital file from being displayed or played by the common viewers or players, and (ii) should not be perceptible by a human eye or ear.

Some similar but not identical requirements could be applied to STL stego channels. In the case of 3D-printed objects, the perceptibility of distortions by a human is rather secondary. Instead, we could require that changes introduced by a sanitizer in the STL file neither (i) affect its printability nor (ii) the quality of the printed part. Both of these constraints require further clarification.

3D printing generally incorporates at least the following two subsequent steps. First, as mentioned in Section~\ref{sec:related_work:STL_stego_channels}, the STL file needs to be sliced into a sequence of layers to be manufactured and to generate a sequence of toolpath commands for each of these layers. Second, these toolpath commands need to be executed by the 3D printer. The success of any of these steps is not necessarily guaranteed.

For example, changes introduced to the location or coordinates, even of a single facet, could break the ``watertight'' model. Should the slicer fail in an attempt to repair such a broken model, no toolpath will be generated. Furthermore, coordinate changes might require the 3D printer to operate outside of the areas where quality is guaranteed\footnote{In the case of Powder Bed Fusion (PBF) 3D printers that are common in metal AM, areas at the edges of build platform often experience higher degree of distortions.} or even outside of the physical dimensions supported by the machine.

For any manufactured object, the term \emph{quality} can be defined as ``fit for purpose,'' meaning that the application area of the part determines the specific requirements. In AM, it is common to break down quality into three F's: Fit, Form, and Function~\cite{gibson2014additive}. Geometry defined in the STL file directly impacts the first two F's and can also impact the last one. For example, geometrical deviations in STL have been used in intentional sabotage attacks~\cite{sturm2014cyber, belikovetsky2017dr0wned} (impact on functional characteristics such as mechanical strength), though not all such modifications will have a negative impact~\cite{yampolskiy2021myths}. For the sake of completeness, we should note that functional characteristics can also be degraded by manipulation of feedstock~\cite{yampolskiy2015security, zeltmann2016manufacturing} or manufacturing process parameters~\cite{yampolskiy2015security, slaughter2017how}.

\paragraph{\textbf{Consequences for Sanitizing:}} 
\textbf{[C4]} 
STL sanitizer should neither (i) affect the design's printability nor (ii) the quality of the printed part. To guarantee this, sanitizer should not ``move'' in 3D space either vertices or facets nor should it introduce new or delete existing facets.

\subsection{Distinguisheability of Stego Data}

The described above presence of multiple stego channels and alternative encoding strategies create an additional challenge -- \emph{distinguisheability} of embedded data. It applies to all stego channels in STL files. It can be subdivided into two problems: (i) identify whether data is embedded in a particular stego channel at all, and (ii) should it be the case, what portion of the file's encoding capacity for this channel is used for the data.

At the point of the writing, it is unclear whether any of these two problems can be solved, and if so, under what conditions. As a rule of thumb, the first problem is commonly resolved by identifying statistical patterns produced by the programs that generate a file with stego channels and then checking whether a specific file violates these patterns.
While we assume that CAD programs that produce STL files would exhibit such patterns as well, e.g., writing out facets semi-sorted, we are unaware of any real studies of such properties. Furthermore, the question remains whether an encoding strategy can be defined that would allow the embedding of data without violating characteristics common for CAD programs.

The second problem is commonly solved similarly. Specifically, one could investigate whether some statistically significant characteristics on one part of the stego channel (which has encoded data) would distinguish it from the characteristics in the rest of the stego channel (where no data has been encoded). However, this approach can be easily defeated. As proposed in the original work by Yampolskiy et al.~\cite{yampolskiy2021what}, a fully encrypted stego channel will use data padding and encryption so that the entirety of a stego channel is overwritten, making this kind of distinguishability impossible. 

\paragraph{\textbf{Consequences for Sanitizing:}} 
\textbf{[C5]} 
Sanitizing needs to be performed regardless of whether it is known (or suspected) that data is embedded in one or several stego channels. Furthermore, the entire encoding capacity of each of the STL stego channels must be sanitized.

\subsection{Preimage Resistance}

While not specific for STL stego channels alone, for completeness, we need to mention the need for \emph{preimage resistance}. 
This means that it should be impossible to restore the originally encoded data based on the sanitized STL file alone.

\paragraph{\textbf{Consequences for Sanitizing:}} 
\textbf{[C6]} 
Sanitizing should be only informed by the encoding capacity in a particular stego channel. The current state of this channel (e.g., bit values in a specific encoding scheme) should not have any effect on the actions taken by the sanitizer.


\section{Sanitizing Stego Channels}
\label{sec:sanitizing}

\begin{figure}[tbp]
    \centering
    \begin{lstlisting}[backgroundcolor = \color{lightgray!10}, frame = single, language = C] % , basicstyle=\tiny

g_NumberRepresentation = STANDARD

func SanitizeSTL (FileName)
{
  carrierSTL = LoadSTL (FileName)

  carrierSTL = SanitizeFacetChannel (carrierSTL)
  carrierSTL = SanitizeVertexChannel (carrierSTL)
  carrierSTL = SanitizeNormalChannel (carrierSTL)

  // No need to explicitly sanitize Number or ASCII stego channels
  // They are "taken care of" by saving STL files uniformly

  SaveSTL (FileName, carrierSTL, g_NumberRepresentation)
}

    \end{lstlisting}
    
    \caption{Pseudo-code for sanitizing all STL stego channels by calling sanitizers for individual channels.}
	\label{fig:code_sanitizing:wrapper}
        
\end{figure}

Figure~\ref{fig:code_sanitizing:wrapper} presents the general approach for sanitizing all stego channels in a specified STL file. Upon loading the STL file into an internal structure $carrierSTL$, the sanitizing function explicitly calls sanitizers for Vertex, Facet, and Normal stego channels. Sanitizing the Number and the non-functional ASCII channels happens implicitly by saving the STL file in a uniform manner. We discuss sanitizers for individual STL stego channels below.

\subsection{Sanitizing Facet Channel}

To sanitize the facet channel, we must randomly rearrange all the facets. As a basis for this, we selected the Fisher-Yates algorithm~\cite{durstenfeld1964algorithm}. The pseudo-code of the adopted algorithm is shown in Figure~\ref{fig:code_sanitizing:facet}.

The algorithm operates on an array of all facets in the design file. In the first iteration, it randomly selects one of the facets and swaps it with the last one in the array. In the second iteration, it randomly selects a facet between the first and the second to last in the array; afterward, this facet is swapped with the second to last one. These steps are repeated similarly until the range from which facets are randomly selected ``shrinks'' to just one. The algorithm ensures that all facets have been moved to random locations in the array.

\begin{figure}[tbp!]
    \centering
    \begin{lstlisting}[backgroundcolor = \color{lightgray!10}, frame = single, language = C] % , basicstyle=\tiny

SanitizeFacetChannel (carrierSTL)
{
    n = len(carrierSTL.facets)
    
    for (i = n - 1; i > 0; i--) {
        // Pick a random index from 0 to i
        j = randint(0, i) 
        
        // Swap facets with indices i and j
        tmp = carrierSTL.facets[i]
        carrierSTL.facets[i] = carrierSTL.facets[j]
        carrierSTL.facets[j] = tmp
    }
  
    return carrierSTL
}
        \end{lstlisting}
    
    \caption{Pseudo-code for sanitizing Facet stego channel (based on Fisher-Yates algorithm~\cite{durstenfeld1964algorithm}). }
	\label{fig:code_sanitizing:facet}
        
\end{figure}




\subsection{Sanitizing Vertex Channel}

In STL format (see summary description in Section~\ref{sec:related_work}), the order in which vertices are listed in the facet's definition determines its outer side per the right-hand rule. 
Therefore, by sanitizing the vertex stego channel, we are limited to the cyclical rotation of the three vertices.
Assuming that the original order is ($v_1$, $v_2$, $v_3$), it can be transformed through rotation to two additional states: ($v_2$, $v_3$, $v_1$) and ($v_3$, $v_1$, $v_1$).

The algorithm to sanitize the vertex channel is shown in Figure~\ref{fig:code_sanitizing:vertex}. 
We randomly select whether the order should be rotated left, rotated right, or remain unchanged.
Note that the algorithm will sanitize regardless of the used encoding strategy. However, if binary encoding is used, this algorithm will disproportionally affect the encoded bit values, depending on the choice made which value is represented by two vertex rotation states and which by just one.

\begin{figure}[h!]
    \centering
    \begin{lstlisting}[backgroundcolor = \color{lightgray!10}, frame = single, language = C] % , basicstyle=\tiny

SanitizeVertexChannel (carrierSTL)
{
    for facet in range carrierSTL.facets {
        turns = ["left", "right", "none"]
        rotation = random.choice(turns)
        switch rotation {
        case "left":
            facet.RotateLeft()
        case "right":
            facet.RotateRight()
        case "none":
            continue
        }   
    }
    return carrierSTL
}
        \end{lstlisting}
    
    \caption{Pseudo-code for sanitizing Vertex stego channel.}
	\label{fig:code_sanitizing:vertex}
        
\end{figure}

\subsection{Sanitizing Normal Channel}
\label{sec:sanitizing:normal}

Sanitizing the normal stego channel is trivial, requiring only re-calculation of the normal vector based on the coordinates of the facet's three vertices (see pseudo-code in Figure~\ref{fig:code_sanitizing:normal}). An additional advantage of this approach is that it ``erases'' even more sophisticated encoding than the binary compliant/non-compliant to the right-hand rule, e.g., use of the normal vector length or direction. 

The only caveat is that this approach implicitly stipulates encoding as a cyclical rotation of the vertices in a facet. 
Should this assumption be incorrect, e.g., encoding allowed ``breaking the order'' could produce results like ($v_2$, $v_1$, $v_3$) instead of the original ($v_1$, $v_2$, $v_3$), the design printability cannot be guaranteed, which would place it out of scope for this work.. 

\begin{figure}[h!]
    \centering
    \begin{lstlisting}[backgroundcolor = \color{lightgray!10}, frame = single, language = C] % , basicstyle=\tiny

SanitizeNormalChannel (carrierSTL)
{
    for each facet in carrierSTL 
      facet.normal = CalculateNormal (facet.v1, facet.v2, facet.v3)

    return carrierSTL
}
        \end{lstlisting}
    
    \caption{Pseudo-code for sanitizing Normal stego channel.}
	\label{fig:code_sanitizing:normal}
        
\end{figure}

\subsection{Sanitizing Number Representation \& Non-Functional ASCII Stego Channels}
\label{sec:sanitizing:number_and_ASCII}

There are two other stego channels that only exist in ASCII STL files. These are the number representation (i.e., scientific vs. standard notation) and classical ASCII file stego channels (e.g., space vs. tabulator). As indicated in the pseudo-code depicted in Figure~\ref{fig:code_sanitizing:wrapper}, neither of these channels requires an explicit sanitizer. Instead, both channels are implicitly sanitized when the geometrical information is re-saved in the STL file in a uniform manner. 


\section{Evaluation}
\label{sec:evaluation}

\begin{figure}[tbp]
    \centering
    \includegraphics[width=0.27\textwidth]{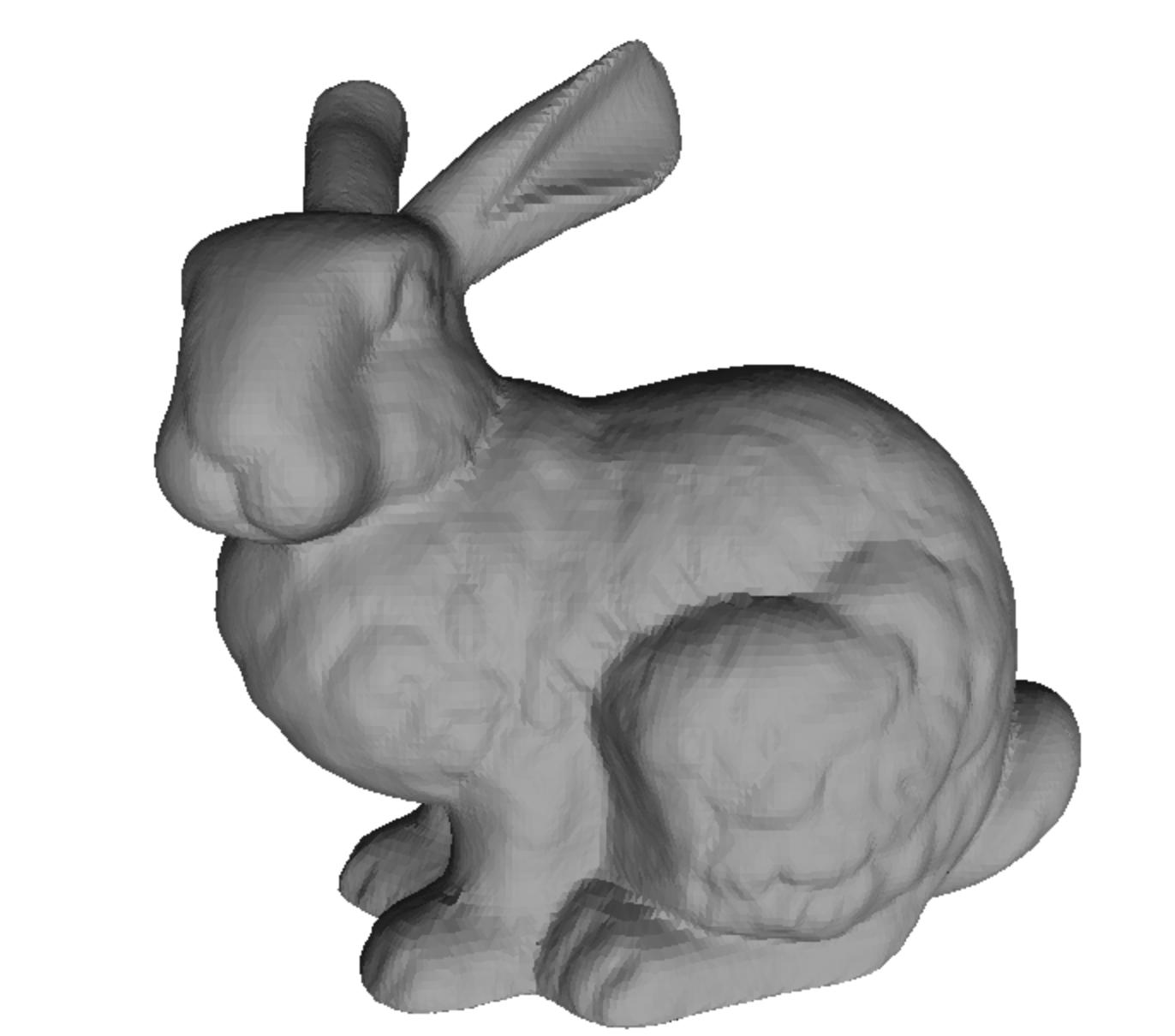}
    \caption{Design used in the evaluation: Stanford bunny~\cite{stanford20143dscanrep}.}
    \label{fig:model_for_evaluation}
\end{figure}

As discussed in Section~\ref{sec:sanitizing}, Normal and ASCII stego channels do not have to be explicitly sanitized, and the result of Normal stego channel sanitization provides consistent results. Therefore, we only evaluated the effect of the developed sanitizer on the remaining vertex and facet stego channels. 
For evaluation, we used a single STL file from the Stanford 3D scanning repository~\cite{stanford20143dscanrep} representing a bunny (see Figure~\ref{fig:model_for_evaluation}). The design features complex geometry common for realistic models. The design's encoding capacity of both Facet and Vertex stego channels is sufficient for storing at least 1024 bits.

Both sanitizers have been tested individually, to assess their effects without interference from other channel's sanitizer. 
For the tests, we generated a random 1024-bit long test sequence. 
We then encoded this test bit sequence into the tested stego channel. Then, the corresponding sanitizer was applied to the file with the encoded information. Finally, we calculated the survivability of individual bits at their specific locations in the STL file. 
To produce statistically significant results, we repeated the encoding and sanitization of the same test sequence 100 times.

\begin{figure}[tbp]
    \begin{subfigure}[b]{0.4\textwidth}
        \centering
		\includegraphics[width=\textwidth]{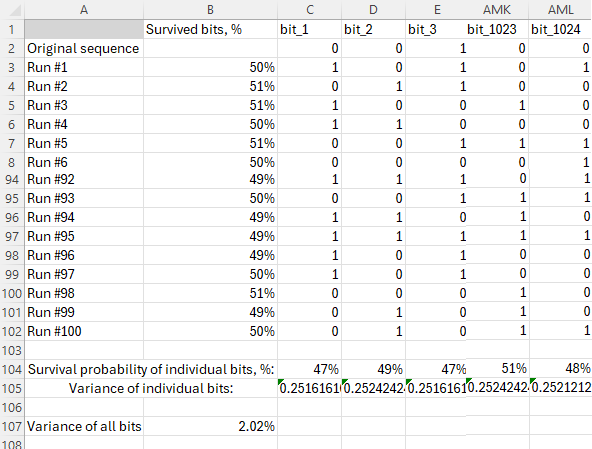}
    
	\caption{Evaluation results (excerpt from CSV)}
	\label{fig:facet_CSV_all_bits}
	\end{subfigure}

    \begin{subfigure}[b]{0.5\textwidth}
        \centering
		\includegraphics[width=\textwidth]{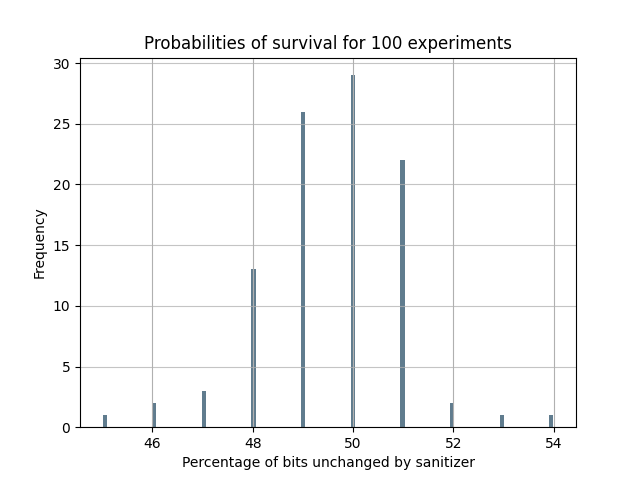}
    
	\caption{Bits unchanged across a test sequence}
	\label{fig:facet_histogram_all_bits}
	\end{subfigure}

    \begin{subfigure}[b]{0.5\textwidth}
        \centering
		  \includegraphics[width=\textwidth]{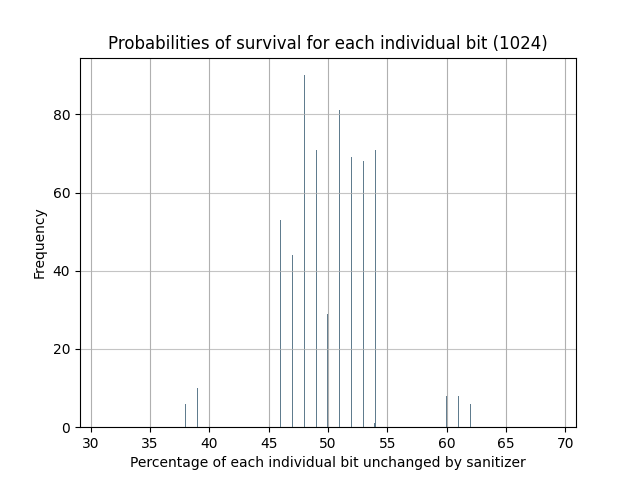}
	    \caption{Individual bits unchanged}
	  \label{fig:facet_histogram_all_individual_bits}
    \end{subfigure}
    
    \caption{Evaluation of facet channel sanitization}
	\label{fig:facet_histogram}
\end{figure}

For the facet stego channel, we used the encoding based on the two consecutive facets. Figure \ref{fig:facet_histogram} shows the empirically determined effects of the sanitizer on the bits in this channel. Figure~\ref{fig:facet_CSV_all_bits} shows an excerpt from the CSV file containing evaluation results. For each test sequence, we recorded which bits have survived sanitizing and which have been modified. Using this data, we calculated an average amount of bits in the test sequence that have ``survived'', i.e., remained unmodified, after sanitizing (see the histogram in Figure~\ref{fig:facet_histogram_all_bits}). On average, this is 50 \% with a variance of 2.02. 
In addition, we analyzed the survivability of each bit in the test sequence. The results of this evaluation are summarized in Figure~\ref{fig:facet_histogram_all_individual_bits}. For each bit, independent of their location in STL file, the average probability of the bit value to remain unchanged is 50 \%. 

\begin{figure}[tbp]
    \begin{subfigure}[b]{0.4\textwidth}
        \centering
		\includegraphics[width=\textwidth]{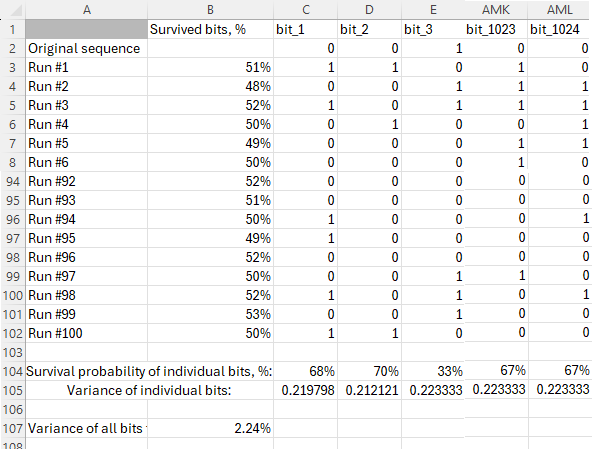}
    
	\caption{Evaluation results (excerpt from CSV)}
	\label{fig:vertex_CSV_all_bits}
	\end{subfigure}

    \begin{subfigure}[b]{0.5\textwidth}
        \centering
		\includegraphics[width=\textwidth]{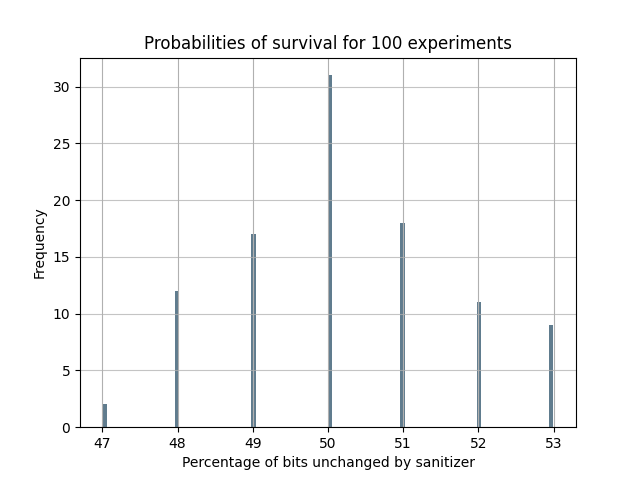}
    
	\caption{Bits unchanged across a test sequence}
	\label{fig:vertex_histogram_all_bits}
	\end{subfigure}

    \begin{subfigure}[b]{0.5\textwidth}
        \centering
		\includegraphics[width=\textwidth]{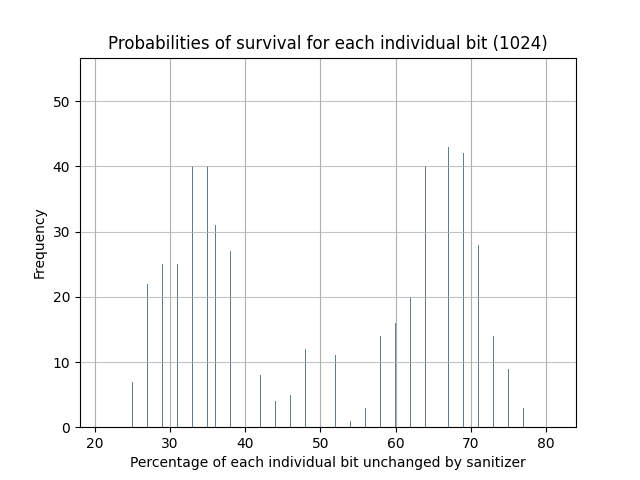}
	\caption{Individual bits unchanged, based on their bit values}
	\label{fig:vertex_histogram_all_individual_bits}
    \end{subfigure}
    \caption{Evaluation of vertex channel sanitization}
    \label{fig:vertex_histogram}
\end{figure}

To evaluate vertex channel sanitization, we used a binary encoding with the ``maximum'' vertex listed first representing bit value 1 and two other variants for bit value 0. Figure~\ref{fig:vertex_CSV_all_bits} shows an excerpt from the CSV file containing evaluation results.
Similarly to the facet channel, we calculated the average amount of bit locations that have not been affected by the sanitizer in the test sequence (i.e., without consideration of the used encoding strategy). 
On average, this is 50 \% with a variance of 2.24 (see the histogram in Figure~\ref{fig:vertex_histogram_all_bits}.
However, when we examine the values of the individual bits under the chosen encoding strategy, the picture changes. 
Instead of a single peak in the middle, we observe two distinct peaks at 33.3 \% and 66.6 \% (see Figure~\ref{fig:vertex_histogram_all_individual_bits}).
This is an expected result that originates from the chosen encoding strategy. 
Encoding a single binary value over a data set with three distinct representations introduces a bias of 1/3 to 2/3 for the selected bit values.

\paragraph{\textbf{Data and Code Release:}}
Working on this research, we developed two open-source libraries.
The first library~\cite{dolgavin2024stlstegorwlib} allows read and write operations on STL stego channels.
The second library~\cite{dolgavin2024stlstegosanitizer} implements the proposed STL stego channel sanitizer.
We release both libraries under the MIT License.
We further release the STL files used and generated during the experimental evaluation and the collected statistics~\cite{dolgavin2024stlsanitizingdata}.


\section{Discussion}
\label{sec:discussion}

While out of scope for this paper, several related aspects must be outlined. Specifically, the existence of the encoding schemes that can be robust against the proposed sanitizer, the distinguishability of encoded information, and the sanitizer's ability to take into account and coexist with legitimate use cases for STL stego channels.

\paragraph{\textbf{Robust Encoding Schemes:}}
In Sections~\ref{sec:challenges:encoding-multitude}, we discussed various encoding schemes that can be defined over STL stego channels.
In Section~\ref{sec:challenges:robust-encoding}, we also demonstrated that some schemes might be robust against naive sanitizing approaches.
While evaluating the proposed sanitizer (see Section~\ref{sec:evaluation}) we empirically demonstrated that all identified STL stego channels will be disturbed; we cannot provide formal proof that no scheme would be robust against the proposed sanitizer.

\paragraph{\textbf{[Minimally] Disruptive Encoding Schemes:}}
In this paper, we intentionally imposed the following two constraints:
(i) the sanitization is performed blindly, i.e., without having access to the original design or its metadata whatsoever;
(ii) the sanitizer uses exclusively transformations that will be non-disruptive, i.e., \emph{neither} interfere with the ability to print a part \emph{nor} alter the design geometry.

However, especially considering the second constraint, disruptive schemes are also possible. 
As outlined in Section~\ref{sec:related_work}, the research literature has extensively studied physical watermarking of 3D printed objects. 
The methods proposed in the watermarking research to embed bit streams by distorting part's exterior or interior geometry could be adopted by malicious actors for various nefarious purposes. 
We also could imagine other schemes that introduce geometrical deviations below 3D printer resolution, thus causing no physical distortion in the printed object (e.g., because coordinates will be rounded to the nearest step position) while providing covert digital channel. 
While the proposed sanitizer does not directly address any of these schemes, it is of an extendible nature. Thus, it is possible to incorporate additional modules as per the specific needs and constraints of the defender.

\paragraph{\textbf{Distinguisheability:}}
Orthogonal to the question of stego channel sanitizing, the question persists whether they contain some information encoded or not. More specifically, a \emph{distinguisher} needs to answer the following questions: (i) Is information encoded in STL stego channels, and if yes, in which ones? and (ii) Which portion of the encoding capacity of the specific channel has been used for this?

Answering these questions would require additional investigations beyond this paper's scope. 
Nevertheless, as an outline of future research, we think that the feasibility to positively answer these questions depends on two factors. 
First, it depends on whether or not the CAD programs that generate STL files are consistent in their behavior.
As programs are commonly deterministic, it is likely --- though not verified yet --- that they will produce statistically significant characteristics. Any deviation from such characteristics can be used as an indicator of embedded information.

Second, the encoding schemes need to be investigated further. 
The question remains whether an encoding scheme can be defined that, even at the expense of encoding rate, could blend in the characteristics of the channels used as generated by legitimate CAD programs.

\paragraph{\textbf{Legitimate Use of Stego Channels:}}
Last but not least, there are legitimate use cases for stego channels. For example, Yampolskiy et al.~\cite{yampolskiy2022crypto} has proposed a scheme for integrating watermarks and fingerprints in STL stego channels to assert IP ownership and enable ``traitor tracing.'' While such schemes have not yet been adopted, it might become necessary to ensure co-existence between a sanitizer and different schemes using stego channels for legitimate purposes.

A possible solution would be to require that legitimate watermarks use a predetermined stego channel and follow a format that could be parsed and supported by the sanitizer. 
For example, a watermark can consist of fields indicating the type of the watermark, the length of the watermark, the certificate ID of the design owner, and the signature of the watermark.
Should the sanitizer be able to verify such a watermark, it could save it before sanitization and be re-applied afterward. 

Furthermore, the sanitizer itself could add information that indicates that the STL file was sanitized. 
The sanitizer must use a predetermined stego channel and encoding strategy to record information in a fixed format.
This information could contain the name of the sanitizer, the date of sanitization, and a signature of this information.


\section{Conclusion}

Additive Manufacturing (a.k.a. 3D Printing) is increasingly adopted throughout manufacturing. Its computerization and reliance on digital design files enabled outsourcing of small-volume contract manufacturing, which became an attractive and widely accepted business model. 
The discovery of the steganographic channels in STL, currently the most widely adopted design file format in AM, could allow malicious actors to piggy-back on the legitimate design files to exfiltrate stolen data from or infiltrate malware into secure environments~\cite{yampolskiy2021what}.

This paper addresses this newly discovered threat category by developing and testing a sanitizer. The proposed sanitizer considers both the characteristics of the STL stego channel and the specific constraints imposed by the use case. Among others, the proposed sanitizer does not affect the ability to manufacture parts using the sanitized design or the quality of the manufactured part. 

In addition to preventing STL stego channels (the job addressed by the sanitizer), the detection of such communication is likely to be of interest to a defender. Therefore, we plan to work on an STL stego channel distinguisher in the future. We are especially interested in answering the following questions:
(i) Is it possible to determine with the high probability that one or multiple STL stego channels have been used to embed additional information? 
(ii) Is it possible to determine the length of this information? 
Answering these questions will require studying statistical characteristics in STL files produced by different CAD programs. It will likely require a more detailed analysis of various encoding strategies available for STL stego channels. 
Until this work on the distinguisher is complete, we strongly believe that the proposed sanitizer can safeguard against unauthorized communication in AM outsourcing models.

\section{ACKNOWLEDGMENTS}
This work was funded in part by the U.S. Department of Commerce, National Institute of Standards and Technology under Grants NIST-70NANB21H121 and NIST-70NANB19H170.
We want to thank Grant Parker for his help in editing this manuscript. We also want to thank the reviewers for their feedback, which helped to improve this paper.

\balance

\bibliography{references}{}
\bibliographystyle{ACM-Reference-Format}


\end{document}